\documentclass[twocolumn,amsmath,amssymb,groupedaddress,prl]{revtex4-1}

\usepackage{graphicx} 
\usepackage{subfigure}
\usepackage{color}
\usepackage{wasysym}
\usepackage{dcolumn}
\usepackage{epstopdf}
\usepackage[]{hyperref}

\hypersetup{
    colorlinks=true,       
    linkcolor=blue,          
    citecolor=blue,        
    filecolor=blue,      
    urlcolor=blue           
}

\begin{document}

\title{Local Mechanical Description of an Elastic Fold}

\author{T. Jules\textsuperscript{1,2}}
\email{theo.jules@lps.ens.fr}
\author{F. Lechenault\textsuperscript{1}}
\author{M. Adda-Bedia\textsuperscript{2}}

\affiliation{\textsuperscript{1}Laboratoire de Physique Statistique, Ecole Normale Sup\'erieure, PSL Research University, Sorbonne University, CNRS, F-75231 Paris, France}
\affiliation{\textsuperscript{2}Universit\'e de Lyon, Ecole Normale Sup\'erieure de Lyon, Universit\'e Claude Bernard, CNRS, Laboratoire de Physique, F-69342 Lyon, France}

\date{\today}

\begin{abstract}
To go beyond the simple model for the fold as two flexible surfaces or faces linked by a crease that behaves as an elastic hinge, we carefully shape and anneal a crease within a polymer sheet and study its mechanical response. First, we carry out an experimental study that consists on recording both the shape of the fold in various loading configurations and the associated force needed to deform it. Then, an elastic model of the fold is built upon a continuous description of both the faces and the crease as a thin sheet with a non flat reference configuration. The comparison between the model and experiments yields the local fold properties and explains the significant differences we observe between tensile and compression regimes. Furthermore, an asymptotic study of the fold deformation enables us to determine the local shape of the crease and identify the origin of its mechanical behaviour. 
\end{abstract}

\maketitle


The process of folding solves the necessity to reduce the space occupied by a large slender object, while keeping the possibility to recover its original shape. Examples of this mechanism are common in nature such as leaves which grow folded inside buds before blooming~\cite{Kobayashi_1998,Grabenweger2005,Couturier2009} or insects that fold their wings inside shells on ground and deploy them on flight~\cite{Forbes1924,Brackenbury_1994,Wootton2003}. To add to their mandatory use, folds allow for shaping on demand three-dimensional structures starting from a two-dimensional thin plate~\cite{Dias2012}. Two extreme examples of such objects are crumpled paper~\cite{Amar1997,Gottesman2018} and origami. The latter is literally the art of folding paper with a specific pattern into sculptures such as the famous paper crane. While origami can be pleasing to the eye, the selection of specific folding patterns generates original properties for the resulting object such as an apparent negative Poisson ratio \cite{Wei2013}. When coupled with the elasticity of the constituent material, origami-based metamaterials are able to generate structures with a wide scope of programmable mechanical properties~\cite{Papa2008,Silverberg2014,Yasuda2015,Chen2016} and shapes~\cite{Dudte2016,Overvelde2016,Dias2014}. The apparent scalability of origami allows for applications that range from small scales with the folding of DNA strands~\cite{Han2017} or tunable microscopic origami machines~\cite{Miskin2018} to large scales with biological systems~\cite{Kuribayashi2006,Mahadevan2005} or the transport of deployable structures in space~\cite{Miura1985}. 

Before studying origami patterns made of a complex network of folds, one should first decipher the behaviour of their most fundamental element: a single fold. Frequently, a fold is defined as two planar surfaces linked by a hinge-like crease setting an angular discontinuity in the structure. From there, different assumptions are made on its properties. If one considers a pure geometrical approach, the faces are assumed to be rigid panels while the crease angle is a degree of freedom of the system. In this case, self-avoidance and kinematic constrains impose relations between the different crease angles of a given origami structure~\cite{Huffman_1976,Wei2013} that constraint the corresponding degrees of freedom. While this model is fine for origami-like systems with rigid structure such as solar panels or dome constructions~\cite{Chudoba2014}, it fails to describe either the deformation of faces in folded membranes~\cite{Papa2008} or the ``snapping" of peculiar bistable origami systems~\cite{Walker2018,Reid2017,Brunck2016,Silverberg2014}. For that, one should take into account both the flexibility of the faces and the mechanical properties of the crease. The former is extensively described by the mechanics of thin elastic sheets~\cite{Audoly2010}. For the latter, experimental work shows a linear response between the moment generated by the crease and its opening angle~\cite{Lechenault2014}. Yet, neither the process by which the crease is created nor the materials properties have been linked to the crease rigidity.

In this paper we will experimentally probe the mechanical response of a single elastic fold. We extend a previous model of the crease acting as a discontinuity between two flexible faces to a new approach in which the crease is described within a continuous fold. This model thoroughly explains the shape of the stressed fold and gives its local mechanical response. We analyse the interesting differences between compressive and tensile loadings of the fold and use the crease reference shape to characterise its rigidity and its linear response. To this purpose, we introduce two length scales $L_c$ and $S_c$ that characterise respectively the spatial extension of the elastic response of the crease and its rigidity.

\section{Experiments}

\begin{figure}[htb]
\begin{center}
\includegraphics[width=0.9\linewidth]{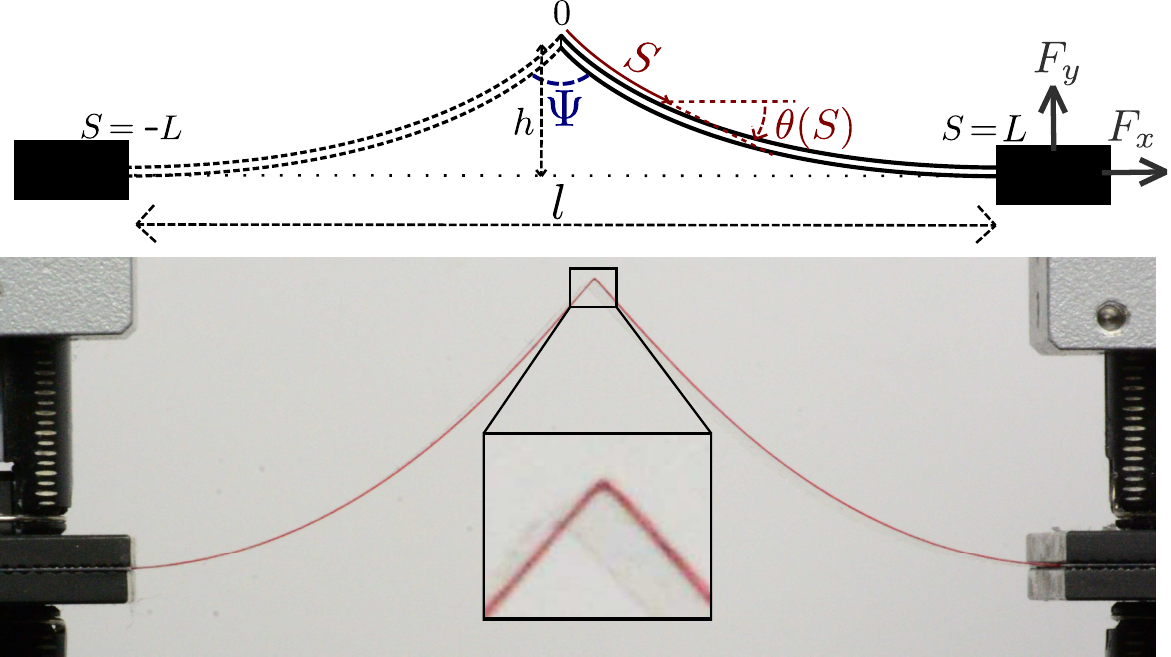}
\caption{Schematics (upper panel) and image (lower panel) of the experimental setup with a zooming of the region around the crease.}
\label{fig:SchemaTheorique}
\end{center}
\end{figure}

We use rectangular flat Polyethylene terephthalate (Mylar) sheets and cut them into strips of length $2L = 10\mathrm{cm}$, width $W = 3\mathrm{cm}$ and thickness $h = 130\mathrm{\mu m}$. We manually pre-crease a strip at half its length and fold it under a $10\mathrm{kg}$ weight for 1 hour. The heavy weight and the plasticity of the material produce the necessary persistent crease of the fold. Then, both free ends of the sample are clamped vertically in an Instron device. The bottom clamp is locked while the upper one is mobile and fixed under a $50\mathrm{N}$ stress gauge which records the vertical applied force $F_x$. To track the shape of the fold, we place a high resolution camera 1 meter away from the device and takes photos of the red coloured side of the fold (see Fig.~\ref{fig:SchemaTheorique}). Both the distance $l$ between clamps, the speed of applied deformation and image acquisition are controlled concomitantly.

\begin{figure}[htb]
\begin{center}
\includegraphics[width=0.9\linewidth]{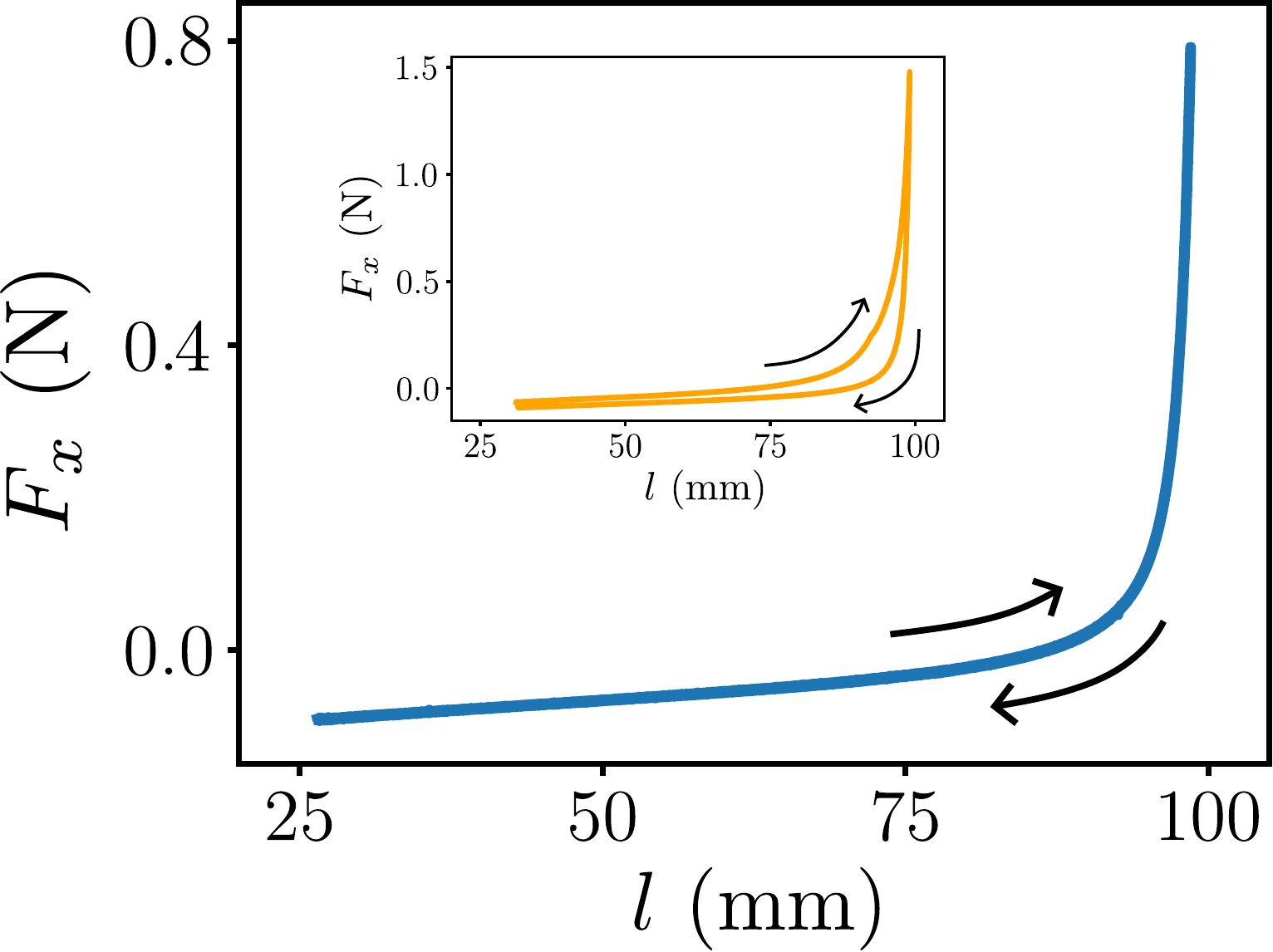}
\caption{Mechanical response of a folded mylar sheet during unfolding-folding cycles. Here, $l$ is the distance between the clamps and $F_x$ is the applied force. The main figure corresponds to an annealed sample and the inset shows the mechanical response of the same sample before annealing. The arrows indicate the direction of the displacement of the upper clamped edge.}
\label{fig:ComparaisonAnnealing}
\end{center}
\end{figure}

A crucial experimental task is the conception of folds with reproducible assigned properties. The process of folding modifies the mechanical properties of the initial sheet because of strain focussing that induces plastic yielding in the vicinity of the creased region. However, for the present study it is desirable that the reference state of the fold does not change during the experiment. To check this, we probe the mechanical response of a folded sample through an opening and closing cycle. Inset of Fig.~\ref{fig:ComparaisonAnnealing} shows that the system is indeed hysteretic. During the cycle, the local stress at the crease is sufficient to cross the plasticity threshold, resulting in a change of the reference configuration~\cite{Benusiglio2012}. Moreover, the creep inherent to the folding process pushes the system at out of equilibrium: it undergoes continuous relaxation~\cite{Thiria2011,Lahini2017}. To take care of both problems, we anneal the fold in a mould that fixes the crease angle (here at $\pi/2$) by first putting the sample in an oven for $45$ minutes at $110^{o}\mathrm{C}$ and then cooling it down at room temperature. This new experimental protocol allows us to recover a reversible mechanical response of the fold (see Fig.~\ref{fig:ComparaisonAnnealing}). Consequently, the annealed sample can be considered as purely elastic and each intermediate state of the system is at mechanical equilibrium.

\section{Fold models}

The mechanical response of the fold, shown in Fig.~\ref{fig:ComparaisonAnnealing}, results from the deformation of both the faces through pure bending and the crease rigidity. To extract the detailed response one should introduce a fold model that separates the contributions of the panels from the creases. In the following we expose two possible models

\subsection{Point-like Crease}

A simple model consists in assuming the fold shown in Fig.~\ref{fig:SchemaTheorique} as two panels of equal length connected by a crease line~\cite{Lechenault2014}. The fold acquires a given equilibrium shape imposed by the combination of applied displacement between clamps and boundary conditions at both the crease and the clamps. The system has a translational symmetry along the fold width, so a two dimensional profile is sufficient to infer the entire shape. The fold is parametrised by the curvilinear coordinate $S$ and the local angle $\theta(S)$ between the tangential vector to the profile line and the horizontal. Here, we define the crease as an elastic hinge between two faces that is solely described by an angular discontinuity $\Psi$ at $S=0$:
\begin{align}
\Psi = \pi + \theta(0^+)-\theta(0^-)\;,
\label{eq:PsiDiscontinu}
\end{align}
with a given energy $U(\Psi)$ per unit length.

To determine the equilibrium shape of the fold, we start by writing the Lagrangian of the system 
\begin{align}
&\begin{aligned}
\mathcal{L} = &\frac{BW}{2L}\int_{-1}^{1} \theta'(s)^{2} \, \mathrm{d}s - F_x L\int_{-1}^{1}\cos \theta(s) \, \mathrm{d}s \\
&-F_y L\int_{-1}^{1}\sin \theta(s) \, \mathrm{d}s + W U(\Psi)\;,
\end{aligned}
\label{eq:lagrangien}
\end{align}
where $s = S/L$ is the dimensionless curvilinear coordinate, $\theta'(s) = \mathrm{d}\theta/\mathrm{d}s$ and $B$ is the bending stiffness of the sheet. The first term is the bending energy of the faces. The second and third terms are the works exerted at the clamps to maintain the imposed distance between clamps and their alignment. $F_x$ and $F_y$ are Lagrangian multipliers associated to the external horizontal and vertical forces applied at the clamps. Finally the last term in Eq.~(\ref{eq:lagrangien}) is the energy of the crease. The minimisation of $\mathcal{L} $ yields the equilibrium Elastica equation satisfied by the fold
\begin{align}
&\theta''(s) - \alpha \sin\theta(s) + \beta \cos\theta(s) = 0\;, \label{eq:elasticaPoint}
\end{align}
with $\alpha = \cfrac{F_xL^2}{BW}$ and $\beta = \cfrac{F_yL^2}{BW}$. The boundary terms resulting from the minimisation of $\mathcal{L}$ read
\begin{align}
&\cfrac{B}{L}\left([\theta'(s) \delta\theta]_{-1}^{0^-} + [\theta'(s) \delta\theta]_{0^+}^{1}\right) + \cfrac{\mathrm{d}U}{\mathrm{d}\Psi} \left(\delta\theta(0^+) - \delta\theta(0^-)\right) = 0\;. \label{eq:BCPoint}
\end{align}
Recall that the fold is clamped at its boundaries, thus the angles $\theta(\pm1)$ are prescribed. The additional boundary conditions at $s=0$ resulting from Eq.~(\ref{eq:BCPoint}) read
\begin{align}
\cfrac{L}{B}\,\cfrac{\mathrm{d}U}{\mathrm{d}\Psi} = \theta'(0^+)=\theta'(0^-)\;.
\end{align}
When the clamps are perfectly aligned one has $\theta(\pm1)=0$ which, coupled to the condition $\theta'(0^+) = \theta'(0^-)$, gives a solution for $\theta(s)$ that is mirror symmetric with respect to $s=0$. For this case, Eq.~(\ref{eq:elasticaPoint}) allows us to show that $F_y = 0$. Finally, we can link $U(\Psi)$ to the macroscopic moments in our system. Integrating Eq.~(\ref{eq:elasticaPoint}) from $s=0^+$ to $s=1$ gives
\begin{align}
&\cfrac{\mathrm{d}U}{\mathrm{d}\Psi} = \cfrac{L}{W} \, F_xh+ \cfrac{B}{L}\,\theta'(1) \;,\label{eq:ForceAngle}
\end{align}
where $h=\int_1^0 \sin\theta(s) \mathrm{d}s>0$. The energy of the crease balances the moments resulting from the applied force and the bending moment of the faces. Notice that the latter was absent in the analysis of Ref.~\cite{Lechenault2014}, an assumption which is valid for large opening angles of the fold, or equivalently for large applied forces. Only in this case the condition $\theta'(1)\approx0$ holds. However, this approximation is insufficient to describe precisely the case of a compressed fold where the moment at the clamp becomes important.

\subsection{Extended Crease}

\begin{figure}[htb]
\begin{center}
\includegraphics[width=\linewidth]{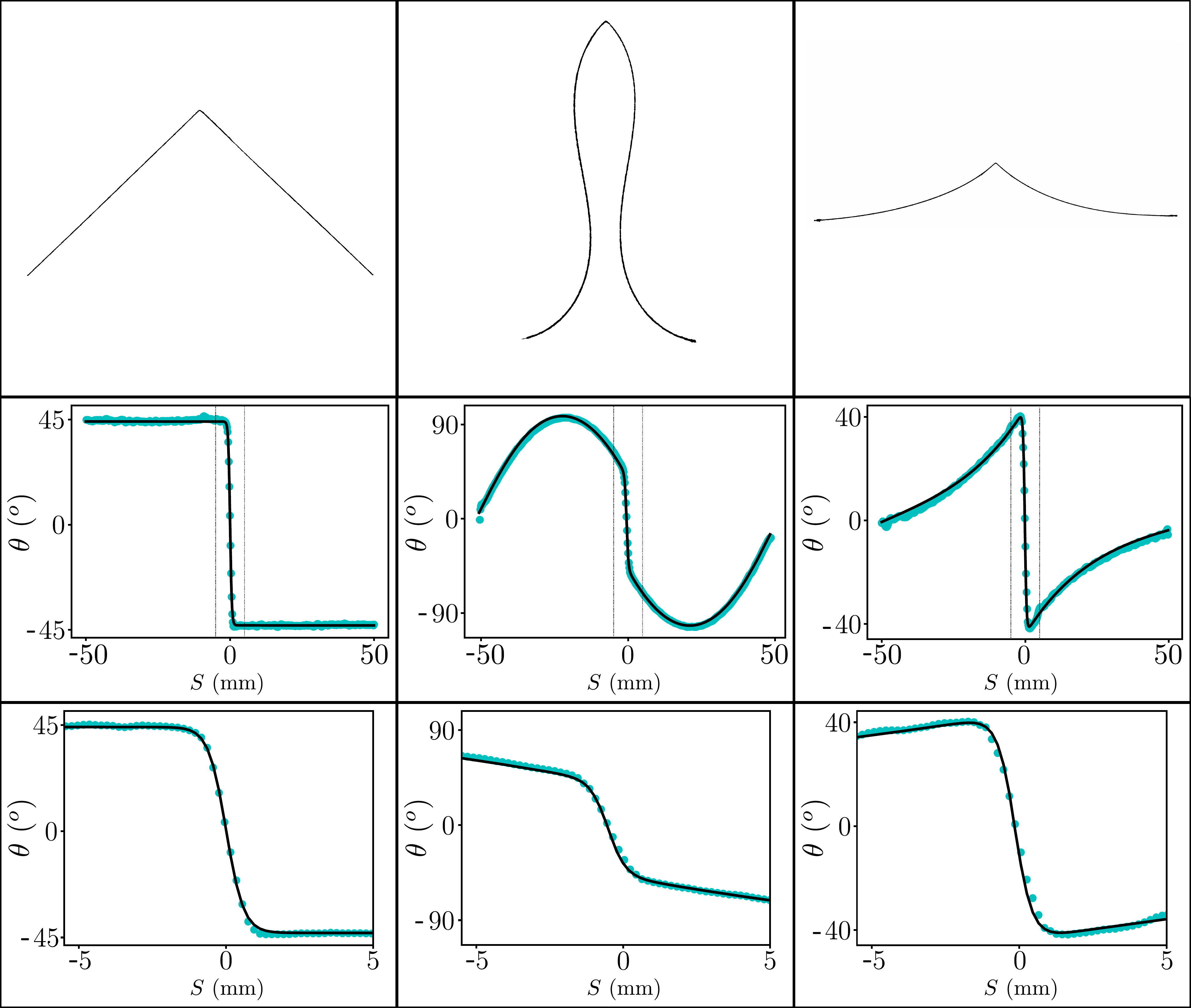}
\caption{Experimental and numerical profile of the fold for three representative configurations. Raw images of the coloured side of the fold are shown in the first row. The second row shows the corresponding evolution of the local angle $\theta$ along the curvilinear coordinate $S$ extracted from experiments (blue points) and computed numerically (solid black line). The last row shows magnifies the same curves around the crease. The first column shows the rest state of the fold. The numerical fitting is made using Eq.~(\ref{eq:rest}) with $\Psi_0 = 92.4^o$ and $S_0 = 0.6\mathrm{mm}$. The second and third columns are respectively examples of a compressed fold with $F_x = -0.127\mathrm{N}$ and $l = 24.5\mathrm{mm}$ and a fold under tension with $F_x = 0.094\mathrm{N}$ and $l = 94\mathrm{mm}$. The computed solutions are found by solving numerically Eq.~(\ref{eq:ElasticaTot}) with $\alpha = -10.5$, $\theta(-1) = 7.1^o$ and $\theta'(-1) = 4.94$ for the compressed fold and $\alpha = 4.65$, $\theta(-1) = -1.4^o$ and $\theta'(-1) =0.44$ for the stretched fold.}
\label{fig:ComparaisonPli}
\end{center}
\end{figure}

As apparent in Fig.~\ref{fig:SchemaTheorique}, our experimental fold is a continuous mechanical system and the crease has a local extension. The description of the crease as a line discontinuity in the local angle might be a reasonable approximation because of separation of scales between the crease extension and system size. Nevertheless, during the crease inception, the system remains continuous and should be considered as so. By construction, the fold is prestressed and its rest configuration is not planar. Therefore, prior to any mechanical testing one should determine the reference configuration of the fold $\theta_0(S)$ that describes the system in the absence of external loading. Fig.~\ref{fig:ComparaisonPli} shows that the reference configuration of the fold is well described by
\begin{align}
\theta_0(S) = \cfrac{\Psi_0-\pi}{2}\tanh\left(\cfrac{S}{S_0}\right)\;,
\label{eq:rest}
\end{align}
where $S_0$ is a characteristic size of the crease region and $\Psi_0$ is the asymptotic angle of the fold. We can envision our system as an elastic plate for which the sheet has a non-Euclidean reference metric $\overline{g}$~\cite{Efrati2013}. The sheet is free of in-plane strains only if its actual metric satisfies $g=\overline{g}$. This solution would correspond to a folded pattern described by the non-uniform field $\theta_0(S)$.

In the presence of external loading, the local elastic deformation of the fold should be defined with respect to the local rest state.
Inspired by the elastic theory of non-Euclidean plates~\cite{Efrati2013}, we postulate that the elastic energy density of a folded sheet is given by $\cfrac{BW}{2}(\theta'(S)-\theta'_0(S))^2$. This expression obviously satisfies a zero energy density for a fold at rest and thus can be interpreted as a bending energy density of a pre-strained elastic plate. To describe the equilibrium state of the fold, we use the same formalism as for the point-like crease but without a separation between the crease region and the panels. We start with a new Lagrangian:
\begin{align}
\begin{aligned}
\mathcal{L} = &\cfrac{BW}{2L}\int_{-1}^{1} \left(\theta'(s)-\theta'_{0}(s)\right)^{2} \, \mathrm{d}s  \\
&- F_x L\int_{-1}^{1}\cos\theta(s)\mathrm{d}s- F_y L\int_{-1}^{1}\sin\theta(s)\mathrm{d}s\;.
\end{aligned}
\end{align}
The minimisation of $\mathcal{L}$ leads to the equilibrium equation and boundary conditions satisfied by the fold shape:
\begin{align}
&\theta''(s)-\theta''_{0}(s) - \alpha \sin\theta(s) + \beta \cos\theta(s) = 0\;, \label{eq:ElasticaTot}\\
&\left[(\theta'(s)-\theta'_0(s)) \delta\theta(s)\right]_{-1}^{+1} = 0\;. \label{eq:BCTot}
\end{align}
For a clamped fold, Eq.~(\ref{eq:BCTot}) is satisfied because the angles $\theta$ are fixed at the boundaries. However, for a free standing fold one should impose $\theta'(\pm1)=\theta'_0(\pm1)$ allowing to recover the rest configuration $\theta(s)=\theta_0(s)$ in the absence of external loading ($\alpha=\beta=0$).

This approach has the advantage to provide us with a continuous description of the fold deformation. In the sequel, we discuss our experimental results within this framework and rationalise the characterisation of the crease.

\section{Results}

Eq.~(\ref{eq:ElasticaTot}) combined with two prescribed conditions at $s=\pm1$ can be solved using classical numerical methods. To take into account unavoidable slight experimental imperfections which are mainly due to misalignment of the clamps, we solve Eq.~(\ref{eq:ElasticaTot}) using an iterative scheme in which not only $\alpha$ and $\beta$ are shooting parameters but also $\theta(-1)$ and $\theta'(-1)$. The four parameters are fixed by fitting the whole shape of the fold to the experimental one. This is done by minimising the sum of the distances between the numerical points and the corresponding experimental ones. The existence and uniqueness of the solution of the Elastica equation~\cite{Pocheau2004} ensure that there is a single set of parameters $(\alpha,\beta,\theta(-1),\theta'(-1))$ for each experimental profile. Recall that if $\theta(\pm1)= 0$ the profile should be mirror symmetric with respect to $s=0$ and consequently $\beta=F_y=0$. Our numerical scheme quantifies correctly the small deviations from perfect alignment of the clamps and gives $\theta(\pm1)\approx 0$ and $|\beta|<<1$ for the whole range of applied deformations.

\begin{figure}[ht]
\begin{center}
\includegraphics[width=0.9\linewidth]{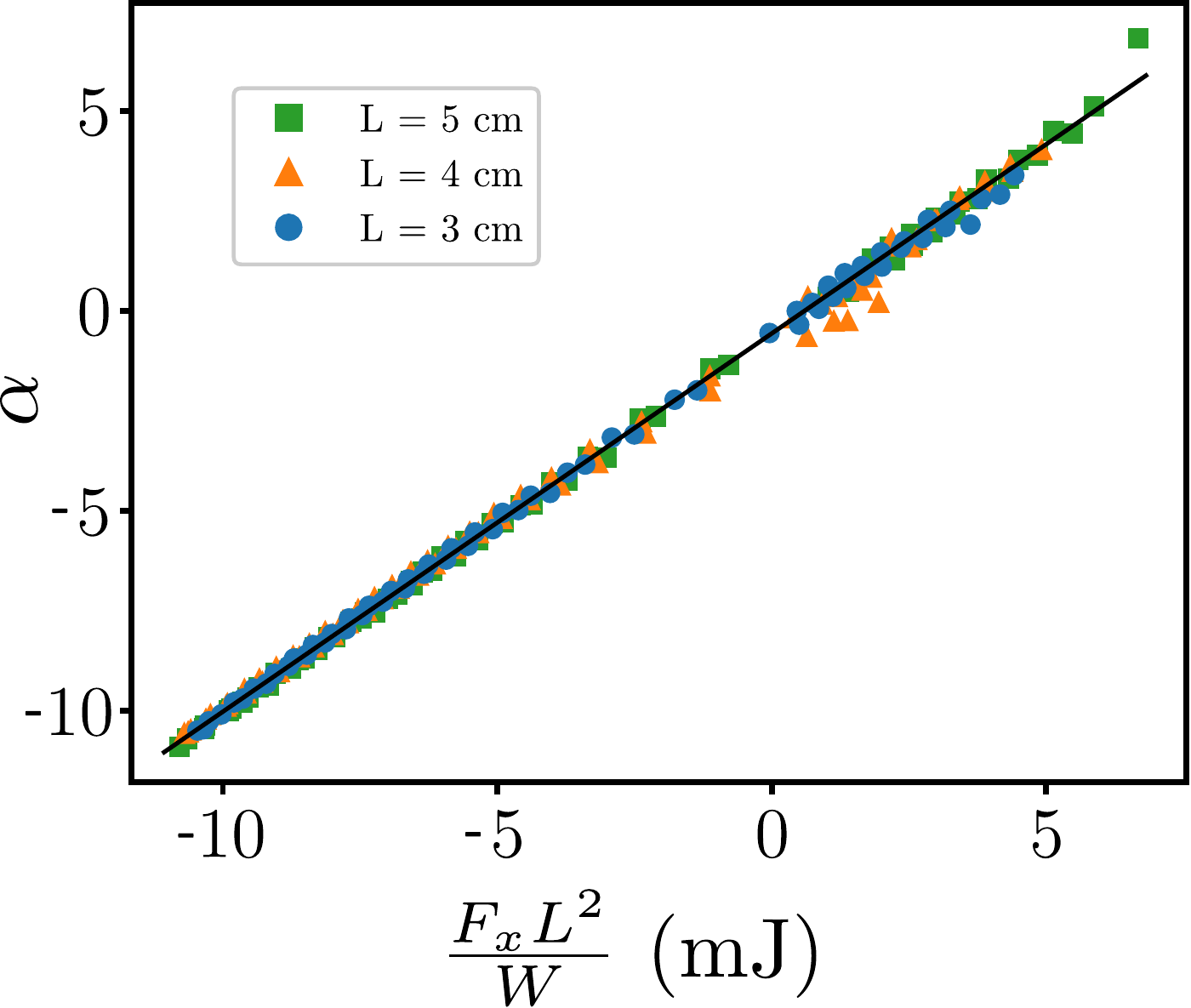}
\caption{Dimensionless parameter $\alpha$, computed numerically, as function of $F_xL^2/W$, measured experimentally. The value of $B = 1.07\mathrm{mJ}$ is inferred from the inverse of the slope of the linear fit represented by the solid line.}
\label{fig:FitB}
\end{center}
\end{figure}

Fig.~\ref{fig:ComparaisonPli} shows a good agreement between theory and experiment regardless of the applied loading. An output of the resolution is the mechanical response of the fold: the parameter $\alpha=F_xL^2/BW$ is determined for each loading test. As a safety test, Fig.~\ref{fig:FitB} shows that the comparison between theoretical results and experiments allows us to recover a constant realistic value of the bending modulus $B$. As shown in the following, the computed profiles serve two purposes. They both assess the robustness of the fold model and contain insights on the local deformation and mechanical response of the creased region. 

\subsection{Compression and extension of the fold}

\begin{figure}[ht]
\begin{center}
\includegraphics[width=0.9\linewidth]{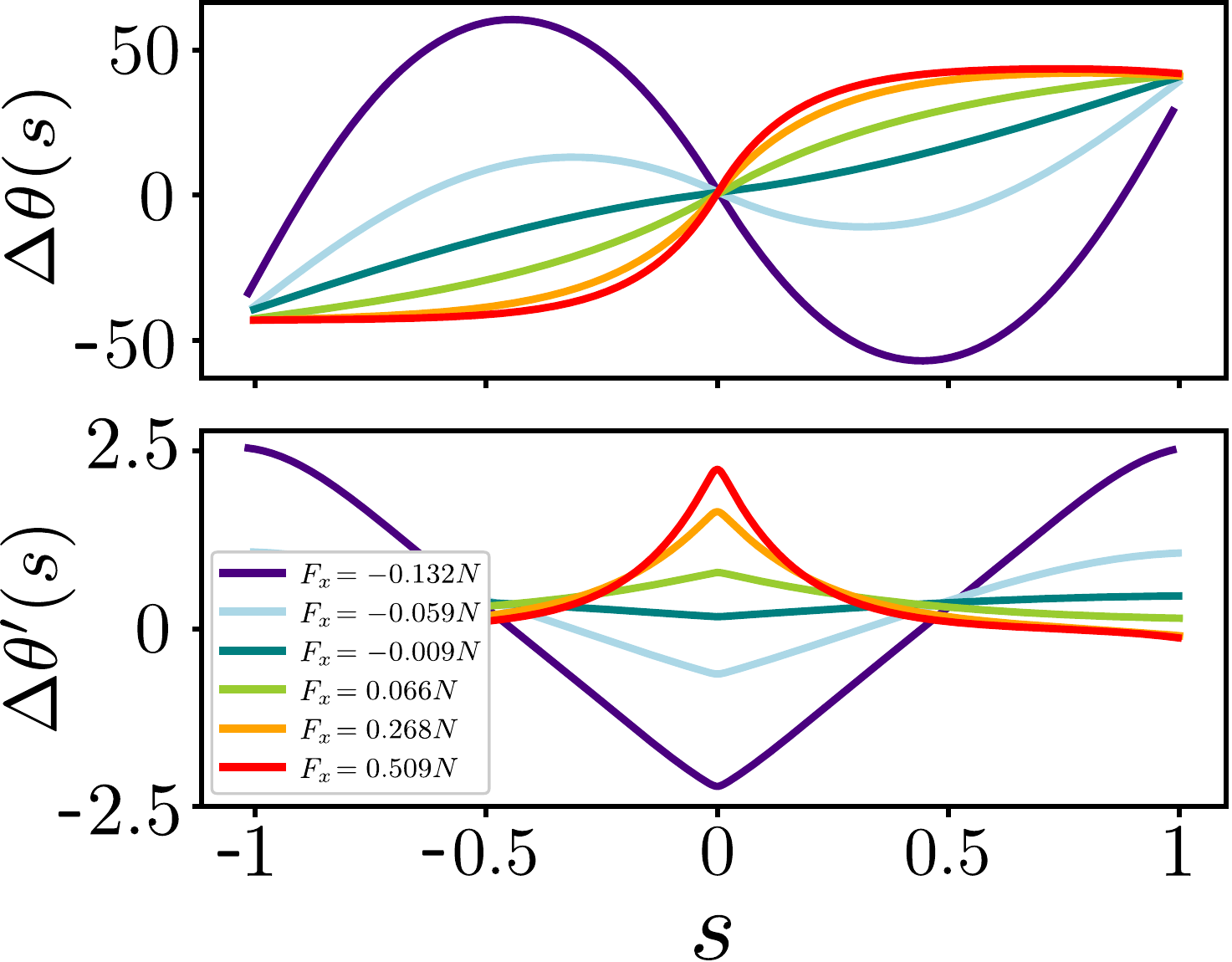}
\caption{Local deformation $\Delta\theta(s)$ and its derivative $\Delta\theta'(s)$ as functions of the normalised curvilinear coordinate. These curves are obtained using the profiles computed numerically for different values of the force $F_x$, or equivalently $\alpha$.}
\label{fig:LocalMomentAngle}
\end{center}
\end{figure}

To analyse the behaviour of the fold in the vicinity of the crease, we introduce the local angular deformation $\Delta\theta(s) = \theta(s)-\theta_0(s)$. We call its derivative $\Delta\theta'(s)$ the local apparent curvature and it represents the local difference between deformed and residual curvature. Fig.~\ref{fig:LocalMomentAngle} shows the evolution of $\Delta\theta(s)$ and $\Delta\theta'(s)$ for different applied forces $F_x$ and allows for interesting observations. First, for $F_x = 0$, the apparent moment, proportional to the apparent curvature, is constant along the fold. This behaviour can be retrieved from the equations of the extended fold model. Plugging $\alpha=\beta=0$ in Eq.~(\ref{eq:ElasticaTot}) and using the boundary conditions $\theta(\pm1)=0$ yields
\begin{align}
\theta(s) = \theta_0(s) -s\, \theta_0(1)\;. 
\label{eq:theta1}
\end{align}
Using Eq.~(\ref{eq:rest}) with $S_0 \ll L$ allows us to write
\begin{align}
\theta(s) = \cfrac{\Psi_0-\pi}{2}\left[\tanh\left(\cfrac{L}{S_0}\,s\right)-s\right]\;.
\label{eq:solf0}
\end{align}
This solution agrees quantitatively both with experiments and the numerics. Therefore, the equilibrium shape given by Eq.~(\ref{eq:solf0}) can be considered as the physical reference configuration of the fold constrained by the experimental situation of Fig.~\ref{fig:SchemaTheorique}. Since no external forces are applied, the moment imposed by the clamping process is completely compensated by the one created by the crease opening.

Eq.~(\ref{eq:solf0}) shows that $\theta'(0)>0$ for $F_x=0$. Therefore, the sign of $F_x$ is insufficient to clearly define whether the crease is opening or closing with respect to its machined residual angle $\Psi_0$. As both the evolution of $F_x$ and the deformation angle are continuous during the experiment (see Fig.~\ref{fig:LocalMomentAngle}), there exists a negative value of the force for which $\theta'(0)=0$.

The local moment also highlights the differences between tensile and compressional regimes. In tension, the kinetic constraints force the faces to be almost flat as it maximises the length along the $x$-axis. The stored elastic energy generated from the work of the external force concentrates on the region where the fold is not flat in the reference state: the crease. Since this region is smaller than the fold length, the deformation quickly becomes harder to increase. This explains why the force increases quickly with $l$ in the tensile regime. Yet, this does not happen in compression, as the geometry allows the faces to bend and store elastic energy. This produces the S-shape often observed with constrained Elastica~\cite{Pocheau2004}. The external work spreads along the whole fold, reaching local extrema for the stress at the crease but also at each clamp. Consequently, if the stress is important enough to reach the plasticity threshold, not only the central crease reference state changes, but a plastic zone can also appear at the clamped edges. 

\subsection{Asymptotic behaviour and crease length}

\begin{figure}[ht]
\begin{center}
\includegraphics[width=0.9\linewidth]{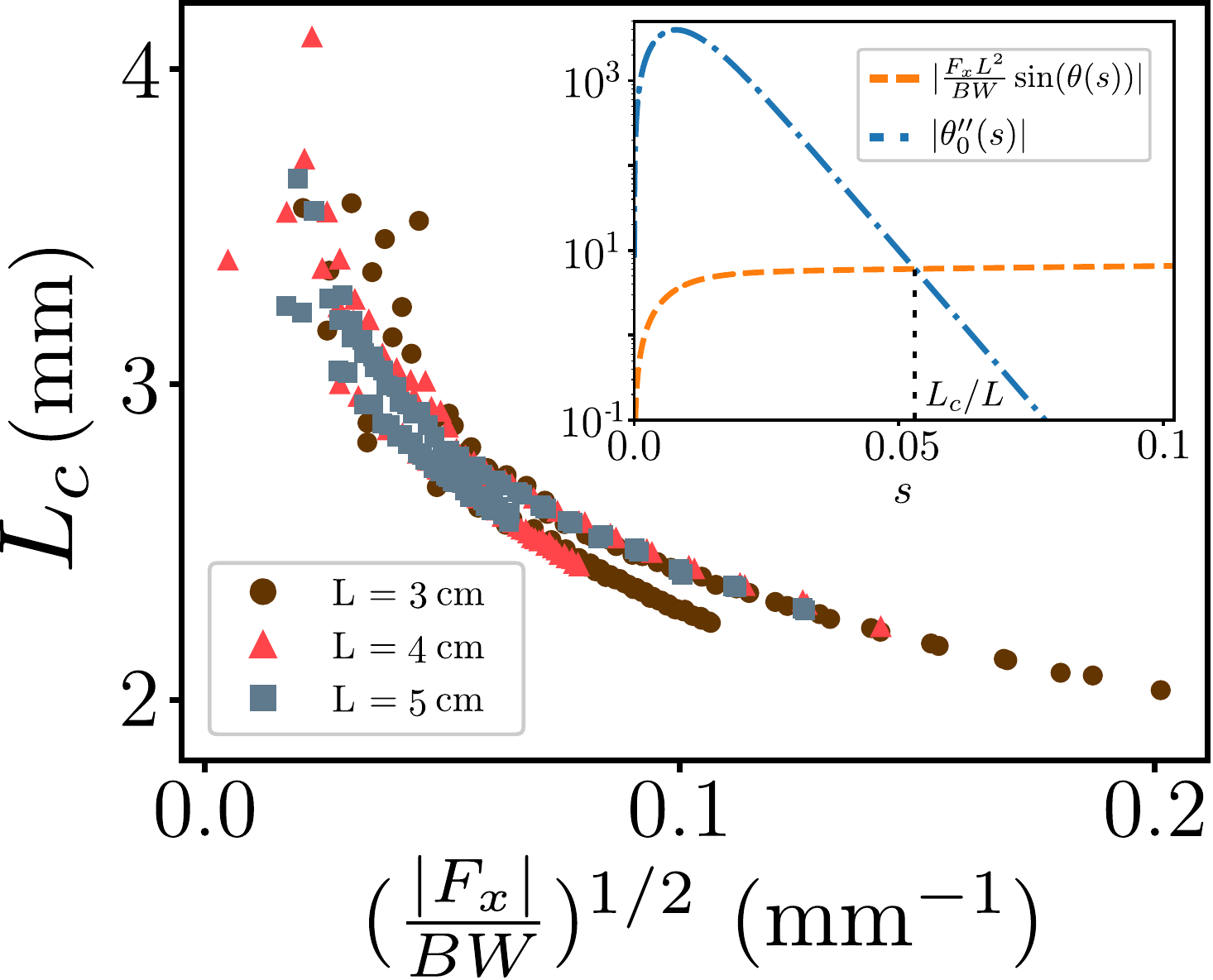}
\caption{Critical length scale $L_c$ as defined by Eq.~(\ref{eq:Lc}) as a function of the inverse length $\left(|F_x|/BW\right)^{1/2}$ for different values of $L$.
Inset: log-lin curve of the absolute value of each term in Eq.~(\ref{eq:ElasticaTot}) with $\beta=0$ near the crease for $F_x =  -0.092\mathrm{N}$.}
\label{fig:EvolLcFinal}
\end{center}
\end{figure}

To link the continuous fold model to the point-like crease, we need to identify the region of the fold that we call the crease, its spatial extension and mechanical properties. The profile $\theta(s)$ of the continuous fold is governed by Eq.~(\ref{eq:ElasticaTot}) which contains two source terms $\theta_0''(s)$  and $\alpha\sin\theta(s)$. They both depend on the position in the fold as shown in the inset of Fig.~\ref{fig:EvolLcFinal}.

The term $\theta_0''(s)$  characterises the local evolution of the residual moment and is crucial to define the extension of the crease as it does not appear in a point-like description. This term dominates Eq.~(\ref{eq:ElasticaTot}) near the fold centre and simplifies it in this region to:
\begin{align}
\theta''(s)-\theta''_0(s) = \Delta\theta''(s) = 0\;.
\label{eq:ElasticaPli}
\end{align}
Since $\theta_0(s)$ is a continuous function, the local angular deformation is simply given by:
\begin{align}
\Delta\theta(s) = T s \;,
\label{eq:NearCrease}
\end{align}
where $T$ is the unique integration constant as the symmetry of the system enforces $\theta(0) = 0$. One meaningful consequence of Eq.~(\ref{eq:NearCrease}) is that the apparent moment in the vicinity of the fold center is constant.

\begin{figure}[htb]
\begin{center}
\includegraphics[width=0.9\linewidth]{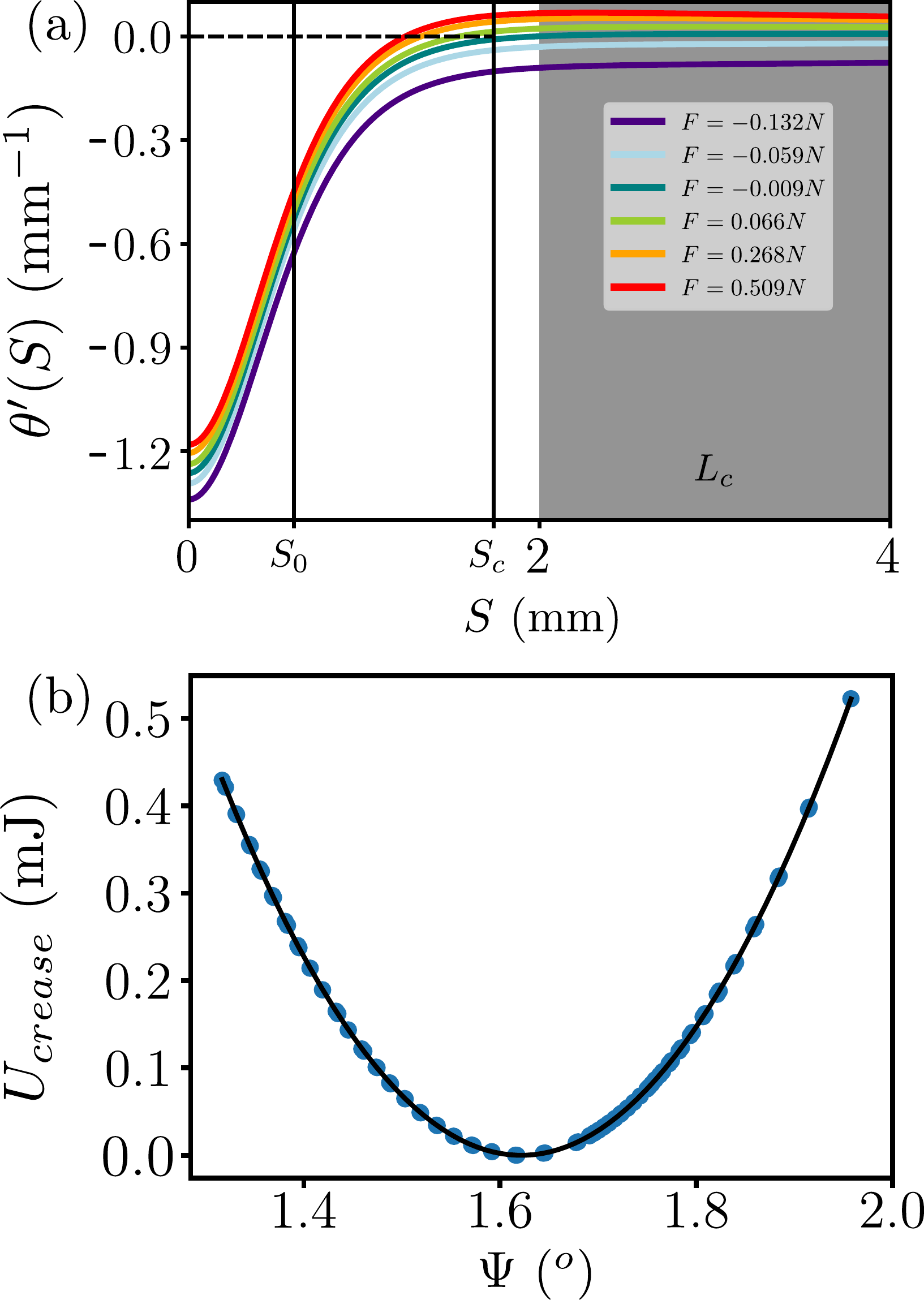}
\caption{(a) Local curvature $\theta'(S)$ near the crease for different external loadings. The three characteristic lengths $S_0$, $S_c$ and $L_c$ are shown, the shaded area corresponds to the region of variation of $L_c$. (b) Bending energy of the crease $U_{crease}$, as defined by Eq.~(\ref{eq:EnergyCrease}), as a function of the crease opening $\Psi$ as defined by Eq.~(\ref{eq:PsiCont}). Both quantities are computed from the numerical fitting of the experimental shape and each blue point corresponds to one photo. The solid line is a quadratic fit $U_{crease} = \kappa W(\Psi - \Psi_0)^2/2$ with $\Psi_0=\pi+2\theta_0(S_c) = 92.9^{o} = 1.62\text{rad}$ and $\kappa=B/2S_c= 308\text{mJ}.\text{m}^{-1}.\text{rad}^{-2}$.}
\label{fig:Sc}
\end{center}
\end{figure}

Far from this region, the second term which represents the action of the external force becomes dominant. In this case, Eq.~(\ref{eq:ElasticaTot}) is simplified into
\begin{align}
\theta''(s) - \alpha\sin\theta(s) = 0\;.
\end{align}
This is the classical Elastica equation used for the discontinuous crease approach to describe bending of the panels. The cross-over between these asymptotic regimes occurs near $s = L_c/L$, where $L_c$ is a length scale that satisfies the condition
\begin{align}
 \theta_0''(L_{c}/L) = |\alpha| \sin \theta(L_{c}/L)\;.
 \label{eq:Lc}
\end{align}
Fig.~\ref{fig:EvolLcFinal} shows that the dependency on the external constraint through $\alpha$ makes $L_c$ largely varying with loading. This confirms that $L_c$ is a poor candidate to characterise the length scale of the crease. The only remaining length scale in this problem is the characteristic length $S_0$ introduced by the rest shape of the fold. As shown in Fig.~\ref{fig:Sc}, $S_0$ gives the right order of magnitude for the extension of the crease. However, it does not separate correctly the crease from the faces, as it is still in the middle of a region where the local moment is still highly varying. To do better, we focus on the physical reference state of the fold ascribed by the experimental configuration of Fig.~\ref{fig:SchemaTheorique}. In that case, the system reaches its minimal elastic energy when $F_x=0$ and the shape of the fold is given by Eq.~(\ref{eq:solf0}). We propose to define the extension of the crease $S_c$ such that the moment at its endpoints vanishes ($\theta'(\pm S_c)=0$). Using Eq.~(\ref{eq:solf0}) and the assumption that $S_0 \ll L$, one finds
\begin{align}
s_c \equiv \cfrac{S_c}{L}= \cfrac{S_0}{2L}\ln\left(\cfrac{4L}{S_0}\right)\;.
\label{eq:sc}
\end{align}
Eq.~(\ref{eq:sc}) shows that $S_c/S_0$ exhibits a weak logarithmic dependency on $L/S_0$ such that $S_0<S_c\ll L$.  As the length of the fold is increased, one has $S_c/L\rightarrow 0$ reaching the point-like crease model. While $S_c$ is defined with respect to the physical reference configuration, Fig.~\ref{fig:Sc} shows that its location within the fold is a clear separation point between two regions with different responses for all loading configurations.

Using Eq.~(\ref{eq:PsiDiscontinu}) and the symmetry of our system, it is legitimate to define the opening angle of the crease $\Psi$ as
\begin{align}
\Psi= \pi +2\theta(S_c)\;,
\label{eq:PsiCont}
\end{align}
and the crease rest angle as $\Psi_0 = \pi+2\theta_0(S_c)$. As $S_c>S_0$, the value of $\Psi_0$ remains close to the one defined by Eq.~(\ref{eq:rest}). The bending energy stored in the crease is given by
 \begin{align}
U_{crease} &=  \cfrac{BW}{2}\int_{-S_c}^{S_c}\Delta\theta'(S)^2\mathrm{d}S\;,
\label{eq:EnergyCrease}
\end{align}
Fig.~\ref{fig:Sc}.(b) shows that $U_{crease}$ is indeed quadratic in the angular crease opening. As $S_c < L_c$, one can derive an analytical expression of the bending energy inside the crease. Using Eq.~(\ref{eq:NearCrease}) one shows that
 \begin{align}
\Delta\theta'(S) &= \cfrac{\Delta\theta(S_c)}{S_c}= \cfrac{\Psi-\Psi_0}{2S_c}\;.
\end{align}
Injecting this expression into Eq.~(\ref{eq:EnergyCrease}), one finds the dependency of the crease energy with its angular opening
\begin{align}
U_{crease} = \cfrac{BW}{4S_c}(\Psi-\Psi_0)^2\;.
\end{align}
Therefore the crease rigidity as described in Ref.~\cite{Lechenault2014} is linked to the elastic properties of the material and the spatial extension of the crease through
 \begin{align}
\kappa = \cfrac{B}{2S_c}\;,
\end{align}
which corresponds to an origami length scale $L^* =B/\kappa= 2S_c$~\cite{Lechenault2014}.

\section{Discussion}

We propose a detailed description of a single elastic fold under mechanical solicitation. Experimental profiles are compared to the shape predicted by a continuous elastic model. Then, we use the local properties derived from the model to explain the differences between compression and tension of the fold, the linear relation between crease opening and the corresponding apparent moment, but also generate a characteristic length marking the transition between crease and flat elastic faces. In the case of a fold long enough for this length to be negligible, the extended model is linked to the point-like one and sets the relation between the crease rigidity, the characteristics of the fold at rest and the material elasticity. 

As one can expect the rigidity of the crease scales like the rigidity of the original material. But it also depends on the inverse of the crease length $S_c$, so the sharper the crease the stiffer it behaves. 
The approximation $S_0 \ll L$ allows to define the crease rigidity in our model as it forces $L^* \ll L$. As a result, the moment due to the crease rigidity always bends the long elastic faces.

\begin{figure}[htb]
\begin{center}
\includegraphics[width=0.9\linewidth]{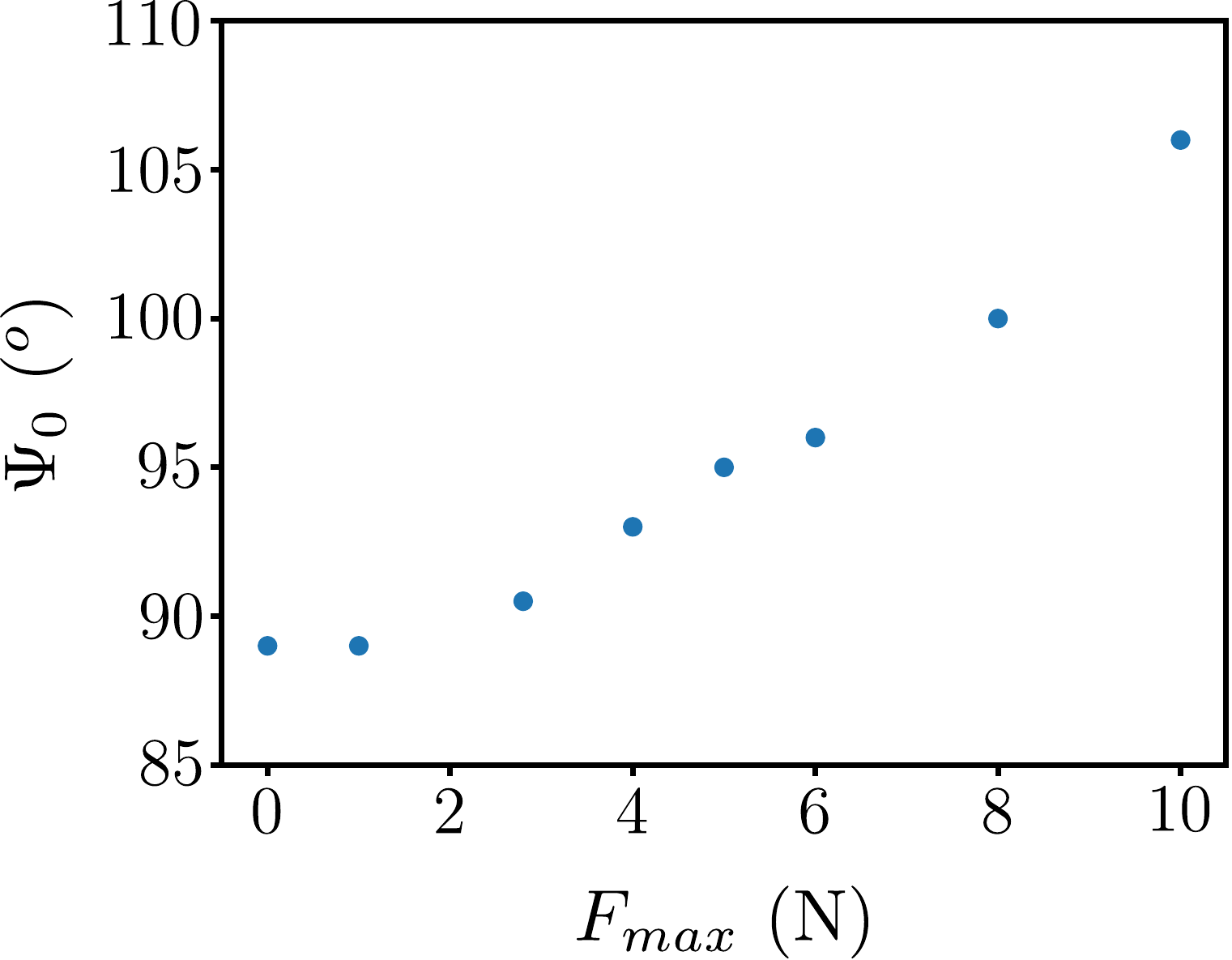}
\caption{Rest angle $\Psi_0$ as function of the maximal vertical force $F_x$ reached while stretching the fold. We increase $l$ until $F_x = F_{max}$ and lock it for one minute. Then, we release the sample and let it relax for one minute before taking a photo of the rest shape to recover $\Psi_0$ as done previously.}
\label{fig:Psi0F}
\end{center}
\end{figure}

This elastic model does not provide a simple scheme to conceive a fold with rigid faces. The only way to soften the crease without changing neither $S_0$ nor the elasticity of the panels is to consider an elastic material with heterogeneous elasticity and a softer crease region. Since $B\propto h^3$~\cite{Audoly2010}, a local reduction of the thickness of the original sheet gives the desired result. Another solution is to change the effective width of the crease by perforation~\cite{Reid2017}. However in both cases, the tempered region becomes plastic for lower stress. It limits the local moment that can be reached~\cite{Benusiglio2012}. If this limit is low enough compared to the moment needed to bend the faces, they remain straight while the crease deforms, though not elastically. Indeed, crossing the plasticity threshold changes the reference configuration of our samples as shown by the evolution of the rest angle $\Psi_0$ in Fig.~\ref{fig:Psi0F}. in high stress situation. 

\begin{figure}[htb]
\begin{center}
\includegraphics[width=0.9\linewidth]{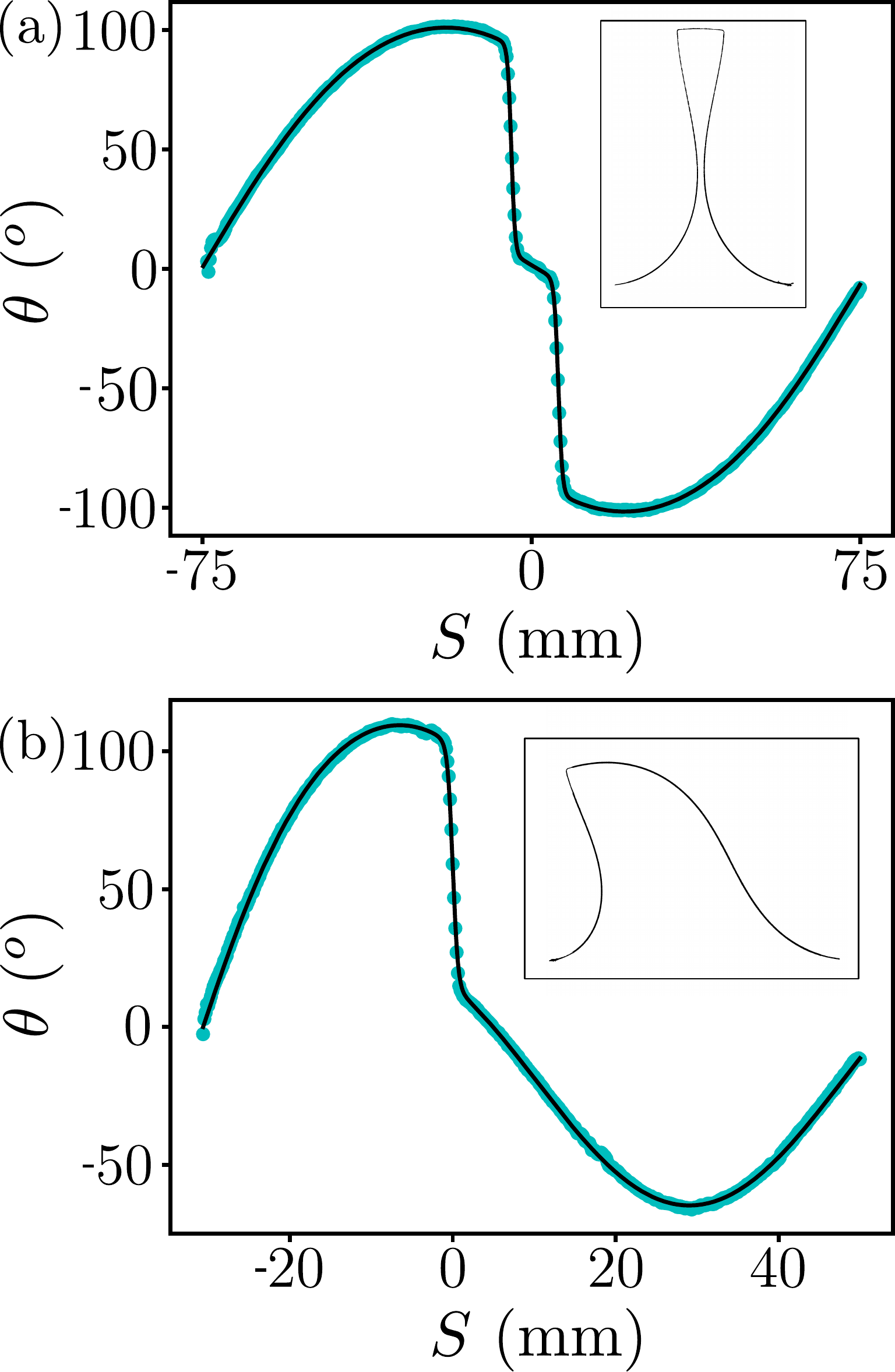}
\caption{Experimental (blue points) and numerical (black line) profile of (a) a compressed fold with two symmetric creases and (b) a compressed asymmetric fold with a single crease. Inset: Black and white images of the samples side.}
\label{fig:DoubleCourbe}
\end{center}
\end{figure}

This model for a symmetric fold with a single crease can easily be extended to more complex structures. In Fig.~\ref{fig:DoubleCourbe}. the numerical solution of Eq.~(\ref{eq:ElasticaTot}) is nicely adjusted to the deformed profile of both a fold with two creases and an asymmetric fold. The main difference comes from the expression of the rest shape $\theta_0$, but in the second case the condition $\beta \approx 0$ is no longer verified as the fold is no longer symmetrical. 

Finally, we have shown that defining the reference configuration of the fold is crucial to understand its mechanical response. The machining of the fold induces residual stresses and defines an absolute rest configuration that can be treated as a non-Euclidian reference metric~\cite{Efrati2013}. However, a fold under loading exhibits a different reference configuration which is the one that imposes the crease extension and not the absolute reference configuration. For complex origami structures where the crease network is interconnected, this underlines the importance of knowing the physical reference configuration by solving the problem in absence of external loading in order to predict the mechanical response of the crease network and to determine correctly the relevant length scales of the problem. Indeed kinematic constraints and mechanical response of the panels, through either bending or stretching, means that the reference configuration of the origami \emph{is not} the sum of the reference configurations of the creases taken independently.

\bibliography{References}
\end{document}